\documentclass[12pt]{iopart}

\usepackage{graphicx}

\begin{document}

\title[Comment to ``On Anisotropic Dark Energy Stars'']
{Comment to ``On Anisotropic Dark Energy Stars''}

\author{Francisco S. N. Lobo%
\footnote[4]{francisco.lobo@port.ac.uk}}
\address{Institute of Cosmology
\& Gravitation,
             University of Portsmouth, Portsmouth PO1 2EG, UK}
\address{Centro de Astronomia e Astrof\'{\i}sica da
Universidade de Lisboa, Campo Grande, Ed. C8 1749-016 Lisboa,
Portugal}

\begin{abstract}

The authors of the paper ``On Anisotropic Dark Energy Stars'',
arXiv:0803.2508 [gr-qc], consider the equations of state
$p_r=\omega \rho$ and $p_t=\omega \rho$, ignoring the fact that
this implies an isotropic pressure, which places strict
restrictions on the values of $\omega$. The authors then argue for
an anisotropic pressure fluid throughout their work, and apply
these equations of state to the energy conditions, consequently
obtaining incorrect intervals for the parameter $\omega$. This
procedure invalidates their criticisms on the following paper:
Class.Quant.Grav. ${\bf 23}$, 1525 (2006).

\end{abstract}

\maketitle


Although evidence for the existence of black holes is very
convincing, a certain amount of scepticism regarding the physical
reality of event horizons is still encountered, and it has been
argued that despite the fact that observational data do indeed
provide strong arguments in favor of event horizons, they cannot
fundamentally prove their existence \cite{AKL}. In part, due to
this scepticism, a new picture for an alternative final state of
gravitational collapse has emerged, where an interior compact
object is matched to an exterior Schwarzschild vacuum spacetime,
at or near where the event horizon is expected to form. Therefore,
these alternative models do not possess a singularity at the
origin and have no event horizon, as its rigid surface is located
at a radius slightly greater than the Schwarzschild radius. In
particular, the gravastar picture, proposed by Mazur and Mottola
\cite{Mazur}, has an effective phase transition at/near where the
event horizon is expected to form, and the interior is replaced by
a de Sitter condensate. The latter is then matched to a thick
layer, with an equation of state given by $p=\rho$, which is in
turn matched to an exterior Schwarzschild solution. It has also
been argued that there is no way of distinguishing a Schwarzschild
black hole from a gravastar from observational data \cite{AKL}.

In Ref. \cite{Lobo:2005uf}, a generalization of the gravastar
picture was explored, by considering a matching of an interior
solution governed by the dark energy equation of state,
$\omega\equiv p/ \rho<-1/3$, to an exterior Schwarzschild vacuum
solution at a junction interface. The motivation for implementing
this generalization arises from the fact that recent observations
have confirmed an accelerated cosmic expansion, for which dark
energy is a possible candidate. Several relativistic dark energy
stellar configurations were analyzed by imposing specific choices
for the mass function. The first case considered was that of a
constant energy density, and the second choice, that of a
monotonic decreasing energy density in the star's interior. The
dynamical stability of the transition layer of these dark energy
stars to linearized spherically symmetric radial perturbations
about static equilibrium solutions was also explored. It was found
that large stability regions exist that are sufficiently close to
where the event horizon is expected to form, so that it was argued
that it would be difficult to distinguish the exterior geometry of
the dark energy stars, analyzed in this work, from an
astrophysical black hole.

The authors of Ref. \cite{Chan}, criticize this work
\cite{Lobo:2005uf} (and other papers, namely, Refs.
\cite{wormhole}) by stating that the procedure for determining the
validity of the interval of the dark energy parameter $\omega$ is
wrong. They then claim to propose a generalization of the limits
of the parameter $\omega$ for the case of ``anisotropic'' fluids
by considering the toy model of a constant energy density,
$\rho=\rho_0>0$, analyzed in Ref. \cite{Lobo:2005uf}. They argue
that special attention must be paid to the limits for the
parameter $\omega$ appearing in the equations of state
$p_r=p_r(\rho)$ or $p_t=p_t(\rho)$, where $\rho$ is the energy
density, $p_r$ the radial pressure and $p_t$ is the tangential
pressure. They further state that a revision is necessary of the
interval of $\omega$, in order to have a correct classification of
dark, standard and phantom energy. Until this point, all is
correct. However, the argument breaks down, when they consider the
equations of state $p_r=\omega \rho$ and $p_t=\omega \rho$ at the
outset, and apply these equations of state to the energy
conditions of an anisotropic fluid. The authors ignore the fact
that taking into account $p_r=\omega \rho$ and $p_t=\omega \rho$
is simply equivalent to considering an isotropic pressure
$p_r=p_t$. One should first and foremost determine the
restrictions imposed on the parameter $\omega$. Thus, due this
flaw in their reasoning, their remaining analysis based on the
bounds placed by the energy conditions is incorrect, as the
authors are effectively considering an isotropic fluid throughout
their work.

Specifically, we shall now go into the mathematical details. Using
the simple toy model of a constant energy density of Ref.
\cite{Lobo:2005uf}, the stress-energy tensor components are given
by
\begin{eqnarray}
\rho&=&\rho_0 \qquad p_r=\omega \rho_0 \,, \\
p_t&=&\omega\rho_0\left[1+\frac{m(r)}{2r}\frac{(1+\omega)(1+3\omega)}
{\omega(1-2m(r)/r)}\right] \,,
\end{eqnarray}
with $m(r)=4\pi \rho_0r^3/3$.

Before going into the analysis of the energy conditions, note that
considering $p_r=p_t$, imposes restrictions on the parameter
$\omega$, given by the following relationship:
\begin{equation}
(1+\omega)(1+3\omega)=0 \,,
\end{equation}
considering $m>0$ and $(1-2m/r)>0$. Thus, the parameter is
restricted to the values $\omega=-1$ or $\omega=-1/3$.

Reviewing the energy conditions, the weak energy condition is
given by $\rho \geq 0$, $\rho+p_r \geq 0$ and $\rho+p_t \geq 0$.
The first condition is readily obeyed as we are considering a
positive energy density, and the remaining two provide the
following inequalities:
\begin{eqnarray}
&&\omega\geq -1 \,, \\
&&3m(r)\omega^2+2r\omega+2r-3m(r) \geq 0 \,,
\end{eqnarray}
respectively.

The authors also consider the strong energy condition given by
$\rho+p_r+2p_t \geq 0$, which provides the following relationship
\begin{eqnarray}
3m(r)\omega^2+4\omega[r-m(r)]+2r-3m(r) \geq 0 \,.
\end{eqnarray}
Note that the respective relationship in \cite{Chan}, namely,
equation $(12)$ is also incorrect.

Now, as the parameter values of $\omega$ are restricted at the
outset by imposing $p_r=\omega \rho$ and $p_t=\omega \rho$ which
implies $p_r=p_t$, the analysis considered by the authors of Ref.
\cite{Chan} on the intervals of $\omega$ is incorrect.

Imposing $\omega=-1$, the above energy conditions provide:
\begin{equation}
\rho+p_r=0, \qquad \rho+p_t=0, \qquad \rho+p_r+2p_t=-2r(1-2m/r)<0
\,.
\end{equation}
Note the violation of the strong energy condition.

Imposing $\omega=-1/3$, from the energy conditions one finally
arrives at
\begin{eqnarray}
\rho+p_r=2\rho_0/3 >0, \qquad \rho+p_t=4r(1-2m/r)/3>0, \\
\rho+p_r+2p_t=2r(1-2m/r)/3>0 \,.
\end{eqnarray}
The strong energy condition is not violated for this case, so that
one does not have a repulsive interior solution, which is a
necessary condition for gravastars or dark energy stars, so that
one may rule out the case of $\omega=-1/3$.

In conclusion, the results of Ref. \cite{Chan} are incorrect,
based on a flaw in the reasoning of the authors from the outset.
Specifically, by considering the equations of state $p_r=\omega
\rho$ and $p_t=\omega \rho$, and ignoring the fact that this
implies an isotropic pressure, the authors argue for an
anisotropic pressure fluid throughout their work, and apply these
equations of state to the energy conditions, obtaining incorrect
intervals for the parameter $\omega$.

\section*{References}



\end{document}